\documentclass[aps,amssymb,twocolumn,prb,subeqn]{revtex4}
\usepackage{bm}
\usepackage{amssymb}
\usepackage{amsmath}
\usepackage{times}
\usepackage{latexsym}
\usepackage{graphicx}
\usepackage{color}
\usepackage{hyperref}

\begin{document}
\title{Current-phase relation for Josephson effect through helical metal}
\author{Christopher T. Olund}
\affiliation{Department of Physics, University of Virginia, Charlottesville, VA 22904}
\author{Erhai Zhao}
\affiliation{School of Physics, Astronomy, and Computational Sciences, George Mason University, Fairfax, VA 22030}
\date{\today}

\begin{abstract}
Josephson junctions fabricated on the surface of three-dimensional topological insulators (TI) show a few unusual properties distinct from conventional Josephson junctions. In these devices, the Josephson coupling and the supercurrent are mediated by helical metal, 
the two-dimensional surface of the TI. A line junction of this kind is known to support Andreev bound states at zero energy for phase bias $\pi$, and consequently the so-called fractional ac Josephson effect. Motivated by recent experiments on TI-based Josephson junctions, here we describe a convenient algorithm to compute the bound state spectrum and the current-phase relation for junctions with finite length and width. We present
analytical results for the bound state spectrum, and discuss the dependence of the current-phase relation on the length and width of the junction, the chemical potential of the helical metal, and temperature. A thorough understanding of the current-phase relation may help in designing topological superconducting qubits and manipulating Majorana fermions.
\end{abstract}
\maketitle

\section{Introduction}
Josephson junctions have been the key elements in superconducting devices such as SQUIDs~\cite{RevModPhys.51.101}.
In the past decades, they have also become the staple components for superconducting
qubits~\cite{RevModPhys.73.357} in the general architecture of circuit quantum electrodynamics~\cite{Schoelkopf:2008fk}. 
Recently it was pointed out that
Josephson junctions patterned on the surface of topological insulators (TI) can be used to create and manipulate
Majorana fermions for topologically protected quantum computation \cite{fu_superconducting_2008}. More generally, TI-based topological qubits
can be integrated with the standard superconducting qubits to achieve a new hybrid platform for information 
processing \cite{PhysRevLett.106.130504,PhysRevLett.106.130505}. These prospects motivate us to carry out a detailed investigation of the equilibrium properties of
TI-based Josephson junctions.

A fundamental property of any Josephson junction is its current-phase relation (CPR), $I(\phi)$, where 
$I$ is the equilibrium supercurrent through the junction and $\phi$ is the superconducting phase 
difference across the junction \cite{RevModPhys.76.411}. For the well known tunnel junction originally considered by Josephson, the
current-phase relation is simply $I(\phi)=I_c\sin\phi$, with $I_c$ being the critical current \cite{Josephson1962251}. By contrast,
the CPR of a pin-hole junction, also commonly referred to as a superconducting constriction, has
a rather different form, $I(\phi)=I_c\sin(\phi/2)$ \cite{ko}. The CPR for junctions between 
unconventional, such as $p$-wave or $d$-wave, superconductors are generally more complicated (for review,
see Ref.~\onlinecite{RevModPhys.76.411}). The CPR is sensitive to the pairing symmetry of the superconductor
as well as the microscopic scattering channels and amplitudes,
and can be measured directly in experiments. 
Anomalies in the CPR often point to new physics. The CPR also controls	
the dynamic properties of the junction, especially when the phase dynamics is slow compared with
the inverse gap. Understanding the CPR is thus important for designing
superconducting circuits and qubits.

The main problem we address is how to find the CPR for the Josephson effect mediated
by a new state of matter, the helical metal at the surface of three dimensional topological insulators~\cite{hasan_colloquium:_2010, qi_topological_2011}. 
Helical metal, consisting of massless Dirac electrons with spin-momentum locking,
is much more exotic than graphene. It is only ``a quarter of graphene," with an odd number of 
Dirac cones (for simplicity we only consider a single Dirac cone at $k=0$ as found in 
Bi$_2$Se$_3$ and Bi$_2$Te$_3$ \cite{hasan_colloquium:_2010, qi_topological_2011}). 
Microscopically, the supercurrent flow is tied to
the process of Andreev reflection, as in the well known case of the Josephson effect through a two-dimensional electron gas.
However, the Andreev reflection of helical Dirac electrons differs from that of 
conventional electrons with quadratic dispersion. One then expects that new scattering kinematics 
such as specular Andreev reflection, which was discovered in the context of graphene by Beenakker 
\cite{PhysRevLett.97.067007}, 
will strongly influence the Andreev bound state spectrum and consequently the CPR in certain
regimes. For example, we will present an interesting scaling relation between the critical current and 
the length of the junction which is unique to helical metal with chemical potential right at the Dirac point. 
In this case, the supercurrent may be thought as being carried by evanescent waves, but it does not decay 
exponentially with the length of the junction.

Many of the new features of the Josephson effect through helical metal were recognized and
discussed in the pioneer work of Fu and Kane~\cite{fu_superconducting_2008}. Most notably, they discovered that a short line
junction at phase bias $\pi$ features a linearly dispersing Andreev bound state spectrum with
a robust crossing at zero energy for transverse momentum $k_y=0$, so the line junction is a ``Majorana quantum wire"~\cite{fu_superconducting_2008}.
The objective of this paper is to generalize their analysis to junctions with finite length
and width, and systematically investigate the effects of finite chemical potential of 
the helical metal and temperature. 
The motivation is to make predictions that can directly compare with experiments. 
Finding the CPR for such junctions turns out to be algebraically cumbersome.
We outline a procedure that is conceptually simple while straightforward to implement.
This also enables us to find a few new analytical results for finite size junctions.

Several groups have successfully fabricated Josephson junctions of various
lengths on exfoliated flakes or epitaxial thin films of Bi$_2$Se$_3$ and observed supercurrent~\cite{SacACpAC:2011vn,
PhysRevB.84.165120,
Veldhorst:2012uq,Qu:2012kx,2012arXiv1202.2323W,2012arXiv1206.6178Y}. It remains unclear that
the supercurrent is entirely due to the TI surface states in all these published results, because the TI bulk also
conducts for many samples used in experiments. This problem can, however, be circumvented by applying a back gate \cite{PhysRevLett.106.196801,doi:10.1021/nl202920p,Kim:2012kx},
chemical doping \cite{PhysRevLett.106.196801,Hong:2012uq}, or adopting the new generation of so-called ideal topological insulators \cite{Zhang:2011vn,Kong:2011zr,Arakane:2012ys} where
the chemical potential is tuned inside the bulk gap. Thus, we will focus on the physics
associated with the helical metal, and assume conduction through the bulk has been eliminated
using one such technique.

The current-phase relation for Josephson junctions on the TI surface has been
investigated theoretically with ferromagnets sandwiched between the superconductors
\cite{PhysRevLett.103.107002, PhysRevB.81.184525}, which introduces
an energy gap to the helical metal.  
In Ref.~\onlinecite{2012arXiv1206.3831Y}, the Josephson effect through helical metal was
considered, but the superconductors are assumed to be conventional BCS superconductors, which differ
substantially from the Fu-Kane model~\cite{fu_superconducting_2008} adopted here.
The anomalous Josephson current via a vortex pinned to a hole drilled through a TI slab was studied in 
Ref. \onlinecite{PhysRevLett.106.077003}.

\section{Model Hamiltonian and solution strategy}

\begin{figure}
\includegraphics[width=0.45\textwidth]{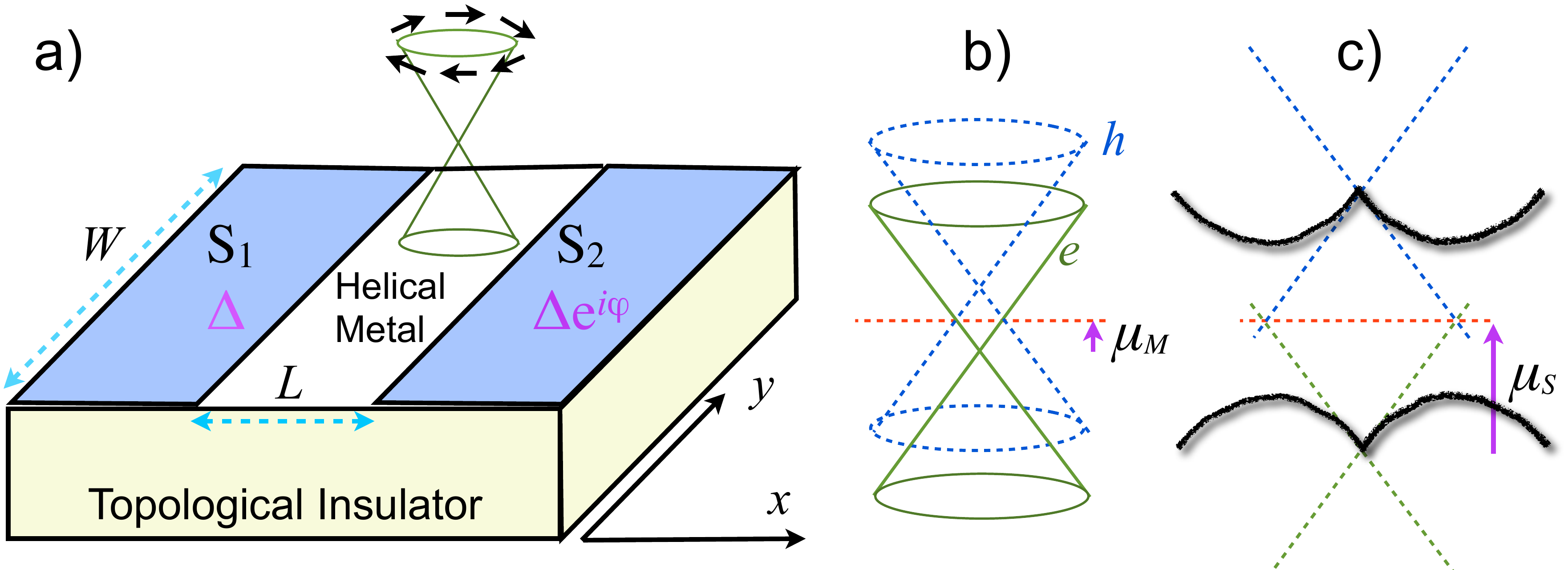}
\caption{(Color online) a) Schematic of a Josephson junction through helical metal. b) The
particle ($e$) and hole ($h$) branch of the excitation spectrum for the helical metal.
c) The gapped spectrum (solid lines) of $H_{S}$ describing the superconductors. The dashed lines show the particle
and hole excitations if the proximity effect is absent ($\Delta=0$). $\mu_M$ and $\mu_S$
are measured from the Dirac point, and are different in general.
}
\label{setup}
\end{figure}

The Josephson junction under consideration is shown schematically in Fig. \ref{setup}. 
The chemical potential of the topological insulator is assumed to be tuned inside the gap
so there is no bulk conduction.
The two-dimensional surface of TI, the helical metal, is modeled by the Hamiltonian \cite{hasan_colloquium:_2010, qi_topological_2011}
\begin{equation}
h_M(\mathbf{k})= v (\sigma_xk_y-\sigma_y k_x) -\mu_{M}.
\end{equation}
Here, $\sigma_{x,y}$ are the Pauli matrices in spin space, $\mathbf{k}=(k_x,k_y)$ is the two-dimensional
surface momentum ($\hbar$ is set to 1 throughout the paper), and $v$ is the velocity of the helical Dirac electrons.
As argued in Ref.~\onlinecite{fu_superconducting_2008}, the presence of an $s$-wave superconductor (S) induces a pairing interaction
between the helical Dirac fermions at the surface of the topological insulator,
and gaps out the surface spectrum. The S-TI interface can then be modeled elegantly by 
 a simple matrix Hamiltonian in Nambu space (we follow the convention of Ref.~\onlinecite{qi_topological_2011}),
\begin{equation}
H_{S}(\mathbf{k})=\left(
\begin{array}{cc}
h_s(\mathbf{k})  &  i\sigma_y  \Delta_s \\
-i\sigma_y \Delta_s^*  &   -h_s^*(-\mathbf{k})
\end{array}\label{fkmodel}
\right),
\end{equation}
where 
\begin{equation}
h_s(\mathbf{k})=v (\sigma_xk_y-\sigma_y k_x) -\mu_S .
\end{equation}
In general, we allow the chemical potential $\mu_S$ and $\mu_{M}$ to be different. For example,
one can add gate control over $\mu_{M}$ in the helical metal region.
The model Eq. (\ref{fkmodel}) has been shown to be accurate at low energies
for both weak and strong coupling between S and TI using self-consistent
calculations \cite{PhysRevB.81.241310,PhysRevB.83.184511,2011arXiv1111.0445G}.

The whole system is translationally invariant in the $y$ direction if the width of the junction is infinite, $W\rightarrow \infty$.
 We define $x=0$ to be on the boundary between the left superconductor S$_1$ and the helical metal, and $x=L$ to be the boundary between the helical metal and the right superconductor S$_2$. The phase of S$_1$ is chosen to be 0, while the phase of S$_2$ is denoted $\phi$. The Hamiltonian is piece-wise constant in each region and has the following generic form:
\begin{equation}
\mathcal{H}(\mathbf{k})=\left( \begin{array}{cccc}
-\mu & v k_+ & 0 & \Delta \\
v k_- & -\mu & -\Delta & 0 \\
0 & -\Delta^* & \mu & v k_- \\
\Delta^* & 0 & v k_+ & \mu
\end{array} \right), \label{gen-ham}
\end{equation}
where  $k_\pm=k_x\pm ik_y$, $\mu=\mu_{M}$ for $0<x<L$ and $\mu_S$ elsewhere, $\Delta(x>L)=\Delta_0 e^{i\phi}$, $\Delta(x<0)=\Delta_0$, and $\Delta(0<x<L)=0$. $\Delta_0$ is the bulk gap of both superconductors. Note that retaining the hole sector
in the helical metal region is crucial for our discussion of the Josephson effect.

To find the CPR, we solve for the Andreev bound state (ABS) spectrum, i.e., solutions to the eigenvalue problem
\begin{equation}\label{eq:Eeig}
\mathcal{H}\psi=E\psi
\end{equation}
with $|E|<\Delta_0$ for given phase difference $\phi$ and $k_y$. While the problem is conceptually
equivalent to finding the bound states in a square potential well,
the algebra is more complicated because of the matrix structure of $\mathcal{H}$.
Our basic strategy is to search for those values of $E$ for which the eigenvector $\psi$
is continuous at $x=0$ and $L$. 

To this end, it is important
to known all the traveling as well as evanescent wave solutions of Eq.~(\ref{eq:Eeig}) in each region.
We use a trick \cite{PhysRevB.82.205331} to organize these solutions which turns out to be crucial in obtaining the ABS spectrum for general parameters.
We first rewrite $\mathcal{H}$ as
\[
\left( \begin{array}{cccc}
0 & 1 & 0 & 0\\
1 & 0 & 0 & 0 \\
0 & 0 & 0 & 1 \\
0 & 0 & 1 & 0
\end{array} \right) vk_x +
\left( \begin{array}{cccc}
-\mu & ivk_y & 0 & \Delta \\
-ivk_y & -\mu & -\Delta & 0 \\
0 & -\Delta^* & \mu & -ivk_y \\
\Delta^* & 0 & ivk_y & \mu
\end{array} \right),
\]
and then rearrange Eq.~(\ref{eq:Eeig}) into an eigenvalue problem for $k_x$,
\begin{equation}\label{eq:keig}
\mathcal{K}\psi=k_x\psi,
\end{equation}
where the matrix $\mathcal{K}$ is non-Hermitian,
\begin{equation}
\mathcal{K}=
\frac{1}{v}\left( \begin{array}{cccc}
ivk_y & E+\mu & \Delta & 0 \\
E+\mu & -ivk_y & 0 & -\Delta \\
-\Delta^* & 0 & -ivk_y &E-\mu \\
0 & \Delta^* & E-\mu & ivk_y
\end{array} \right),
\end{equation}
and the eigenvalue $k_x$ is generally complex.
For given $\phi$, $E$, and $k_y$, we can solve Eq. (\ref{eq:keig})
easily. The complex $k_x$ solutions describe evanescent waves.
Physically, the wavefunctions
for ABS decay inside the superconductor. 
As another example, the bound states for $\mu_{M}=0$
involve evanescent rather than traveling wave solutions of the helical metal Hamiltonian $h_M$ 
(and its counterpart in the hole sector). 

To demonstrate the procedure, we start by discussing the simple case of $k_y=0$ analytically.
This corresponds to Josephson coupling through a TI nanoribbon with small $W$, where transverse quantization
of $k_y$ only allows the single mode $k_y=0$ below the energy scale $\Delta_0$. Note that the length
of the nanoribbon $L$ is kept general.

\subsection{Zero Energy Solution}
We first examine the case of $E=0$ and $k_y=0$. Solving (\ref{eq:keig}), we find eigenvalues of
\begin{subequations}
\begin{align}
        k_x^1=\frac{1}{v}(i\Delta_0+\mu_S)\\
        k_x^2=\frac{1}{v}(i\Delta_0-\mu_S)\\
        k_x^3=\frac{1}{v}(-i\Delta_0+\mu_S)\\
        k_x^4=\frac{1}{v}(-i\Delta_0-\mu_S)
\end{align}
\end{subequations}
with associated eigenvectors (not normalized)
\begin{subequations}
\begin{align}
        \psi_1=(ie^{i\phi},ie^{i\phi},-1,1)^T\\
        \psi_2=(-ie^{i\phi},ie^{i\phi},1,1)^T\\
        \psi_3=(-ie^{i\phi},-ie^{i\phi},-1,1)^T\\
        \psi_4=(ie^{i\phi},-ie^{i\phi},1,1)^T
\end{align}
\end{subequations}
for the superconductor, and degenerate solutions of
\begin{subequations}
\begin{align}
        k_x^1=k_x^2=\frac{1}{v}\mu_{M}\\
        k_x^3=k_x^4=-\frac{1}{v}\mu_{M}
\end{align}
\end{subequations}
with
\begin{subequations}\label{degenevecs}
\begin{align}
        \psi_1=(1,1,0,0)^T\\
        \psi_2=(0,0,-1,1)^T\\
        \psi_3=(-1,1,0,0)^T\\
        \psi_4=(0,0,1,1)^T
\end{align}
\end{subequations}
for the helical metal. The wave function in each region
is a superposition
\begin{equation}
\psi=\displaystyle\sum\limits_{j=1}^4 a_j\psi_je^{ik_x^jx}\;.
\end{equation}
Because the ABS wave function must go to 0 at $x=\pm\infty$, we only include the two eigenfunctions with positive imaginary parts of $k_x$ for S$_2$, and similarly the two solutions with negative imaginary parts for S$_1$. All four eigenfunctions are used for the helical metal.  Requiring continuity of $\psi$ at $x=0$ and $x=L$, we can solve for the eight unknown coefficients. Doing so, it is straightforward to show that the unique solution for zero energy bound states is always at
\begin{equation}
\phi=\pi.
\end{equation}
This agrees with the results of Ref.~\onlinecite{fu_superconducting_2008}.

\subsection{Finite Energy Solutions}

Now we move on to general bound states at finite energy for $k_y=0$. It is convenient to parameterize $E$ using an angle $\beta$, $E=\Delta_0\cos\beta$, for $|E|<\Delta_0$. For the superconductors, we find eigenvalues and eigenvectors of
\begin{subequations}
\begin{align}
        k_x^1=\frac{1}{v}(\mu_{S}+i\Delta_0\sin\beta)\\
        k_x^2=\frac{1}{v}(\mu_{S}-i\Delta_0\sin\beta)\\
        k_x^3=\frac{1}{v}(-\mu_{S}+i\Delta_0\sin\beta)\\
        k_x^4=\frac{1}{v}(-\mu_{S}-i\Delta_0\sin\beta)
\end{align}
\end{subequations}
and
\begin{subequations}
\begin{align}
        \psi_1=(e^{i(\phi+\beta)},e^{i(\phi+\beta)},-1,1)^T\\
        \psi_2=(e^{i(\phi-\beta)},e^{i(\phi-\beta)},-1,1)^T\\
        \psi_3=(e^{i(\phi-\beta)},-e^{i(\phi-\beta)},1,1)^T\\
        \psi_4=(e^{i(\phi+\beta)},-e^{i(\phi+\beta)},1,1)^T
\end{align}
\end{subequations}
In the helical metal region, we have eigenvalues of
\begin{subequations}
\begin{align}
        k_x^1=\frac{1}{v}(\mu_{M}+\Delta_0\cos\beta)\\
        k_x^2=\frac{1}{v}(\mu_{M}-\Delta_0\cos\beta)\\
        k_x^3=\frac{1}{v}(-\mu_{M}-\Delta_0\cos\beta)\\
        k_x^4=\frac{1}{v}(-\mu_{M}+\Delta_0\cos\beta)
\end{align}
\end{subequations}
and the same eigenvectors as in Eqs.~(\ref{degenevecs}).

By matching the wavefunctions at the two boundaries, we find that the bound state energy $E$ has to satisfy the following transcendental equation,
\begin{equation}
E=\pm \Delta_0\cos [\frac{EL}{\hbar v}\pm \frac{\phi}{2}].
\label{transcen}
\end{equation}
This analytical result demonstrates a remarkable feature of these junctions:
the ABS energy does not depend on $\mu_M$ or $\mu_S$ (this is only valid for $k_y=0$).
As a sanity check, for $E=0$ the solution is $\phi=\pm\pi$ which we have found earlier:
zero energy states are always at $\phi=\pm\pi$.
In the short junction limit, $L\rightarrow 0$, we have 
\begin{equation}
E=\pm \Delta_0\cos(\frac{\phi}{2}).
\end{equation}
which was obtained by Fu and Kane earlier in Ref.~\onlinecite{fu_superconducting_2008}. For long junctions, 
there are many quantized ABS levels which only slowly disperse with $\phi$,
\begin{equation}
E_n\sim \frac{n\pi }{2}\frac{\hbar v}{L},
\end{equation}
where $n$ is an integer.

\section{Andreev bound states}

Now we describe how the ABS spectrum can be obtained numerically for the general $k_y$. 
Continuity of each wave function component at $x=0$ and $x=L$ gives in total eight equations.
These boundary conditions can be organized neatly into a matrix
equation in the form of $Az=0$, where the $8 \times 8$ matrix $A$ is a function of $E$, $\phi$, and $k_y$, and 
$z$ is a column vector
containing the eight unknown coefficients. A nontrivial solution requires the determinant of matrix $A$, $D=\det A$, to be 0. Then to find the allowed energies $E$ for a given $k_y$ and $\phi$, we just have to find the zeroes of $D(E)$ in the range $-\Delta_0<E<\Delta_0$. Given the particle-hole symmetry of the problem, we only need to look in the range $0\le E<\Delta_0$. $D(E)$ is in general complex, so we look for zeroes of its absolute value. Numerically it is much easier to search for minima of $|D(E)|$ rather than the zeroes directly. Therefore,
our algorithm starts by splitting the energy range $0\le E<\Delta_0$ into $N$ equal slices. Within each of these slices, we perform a standard golden section search for minima, as described in chapter 10 of Numerical Recipes \cite{nr}.
After a point is identified to be a local minimum, we check if $|D(E)|$ at that point is close to zero (less than a tiny error tolerance). We also check the endpoints of the slices to see if they are zeroes. $N$ has to be sufficiently large to exhaust all the zeros. Finally we check and eliminate unphysical solutions, e.g., those leading to a matrix $A$ of rank lower than 8.

\begin{figure}[h]
\includegraphics[width=0.45\textwidth]{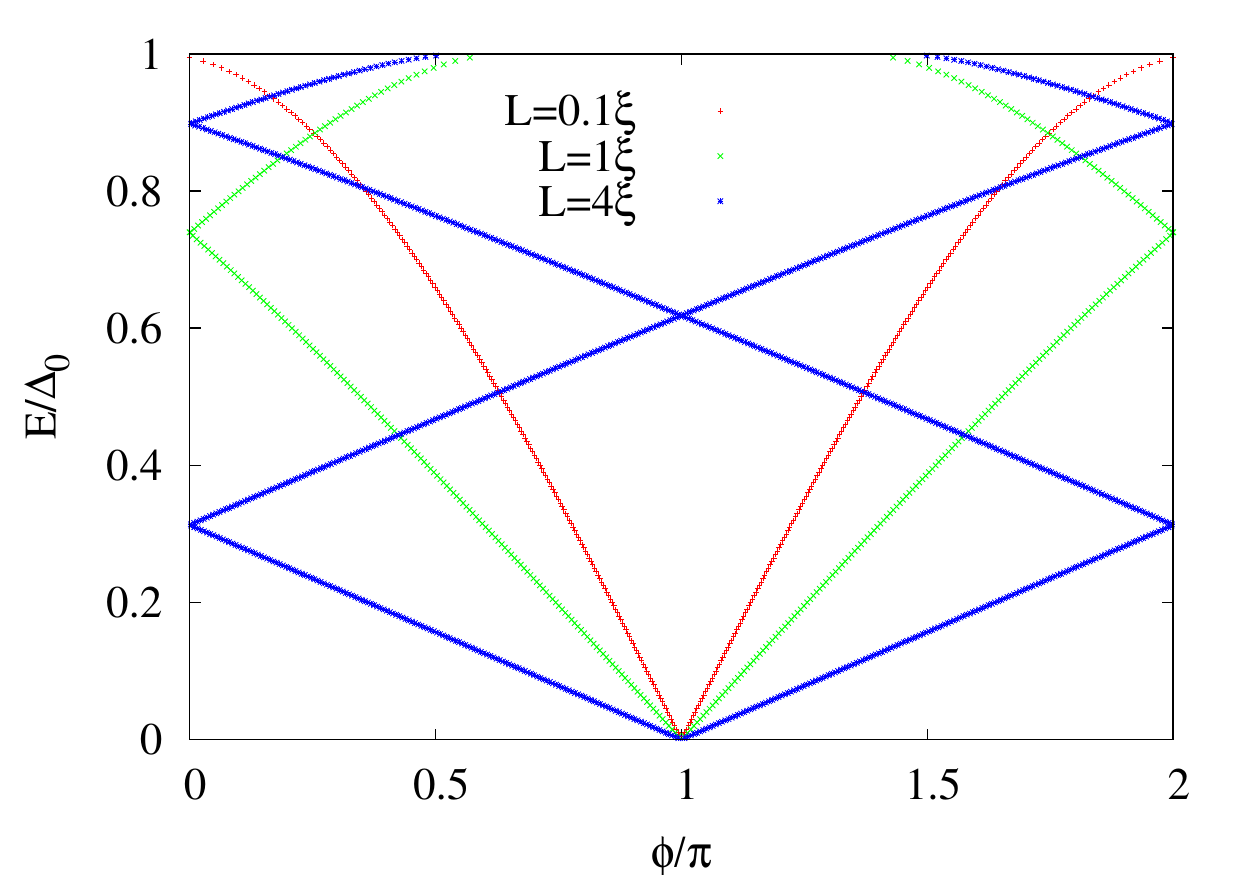}
\includegraphics[width=0.45\textwidth]{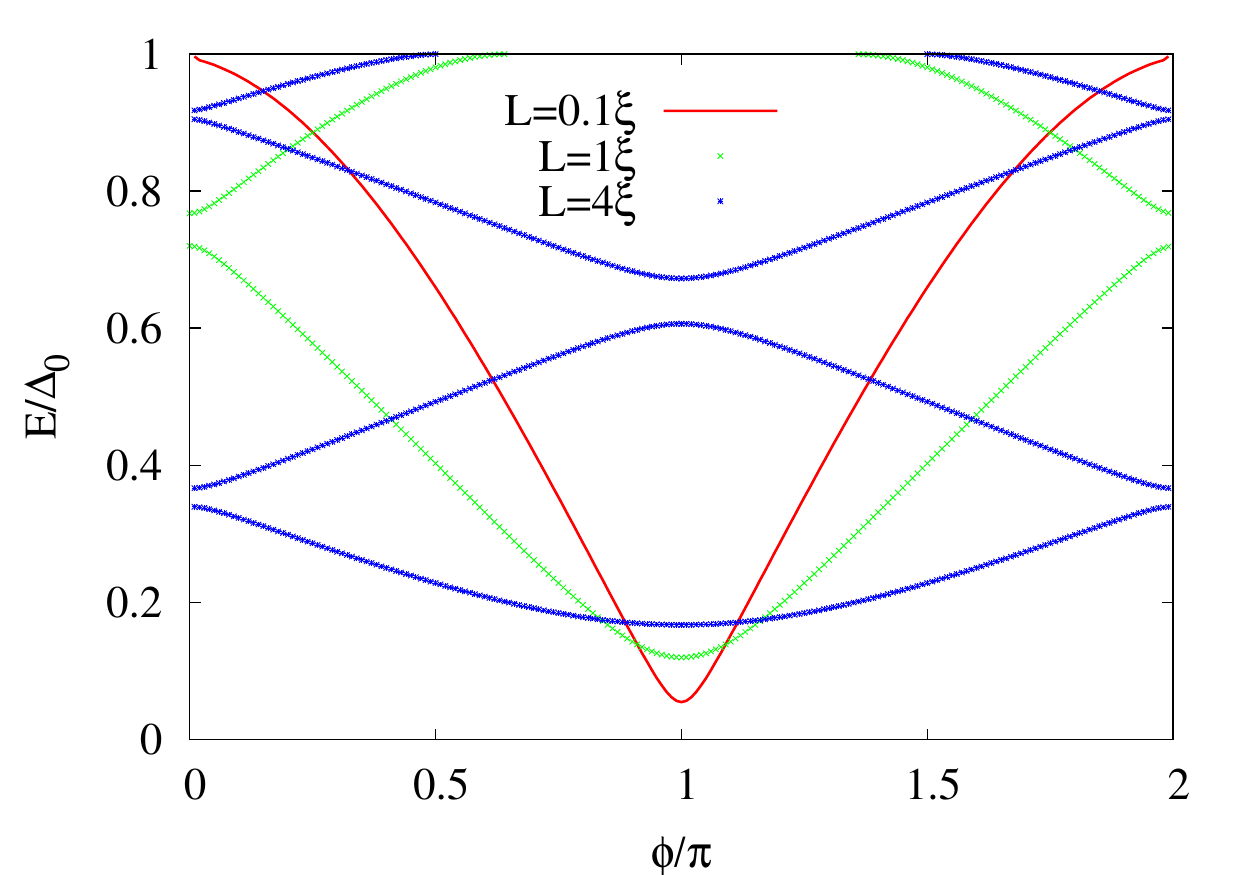}
\caption{Andreev bound state energies as functions of the phase difference $\phi$. 
Only $E\geq 0$ are shown. 
Top panel: $k_y=0$. Bottom panel: $k_y=0.2\xi^{-1}$.
The junction parameters are $\mu_S=2\Delta_0$, $\mu_M=0$, $W=\infty$, and 
the junction length $L$ is measured in units of $\xi=\hbar v/\Delta_0$. 
}\label{abs-plot}
\end{figure}

Fig.~\ref{abs-plot} compares the ABS spectrum $\{E_n(\phi)\}$ for $k_y=0$ and $k_y=0.2\xi^{-1}$,
where the ``coherence length" $\xi=\hbar v/\Delta_0$, and $\mu_M=0$. Here we assume the junction is infinite
in the $y$ direction and $k_y$ is a good quantum number. In each case, we see that as
the junction length $L$ is increased, more branches of ABS show up. The key difference is that
in the case of finite $k_y$, the crossings at $\phi=0$ and $\pi$ observed for $k_x=0$ are now 
replaced by avoided crossings, and $E_n(\phi)$ become smooth functions.

We notice two singular features in the function $E_n(\phi)$.
The first is the crossing at zero energy and $\phi=\pi$. Strictly speaking, the crossing is
only for $k_y=0$. But for small $k_y$, the change in the slope of $E_n(\phi)$ is still 
rapid, and even for large values of $k_y$, the slope of $E_n(\phi)$ changes sign at $\phi=\pi$.
The second is the merging of the ABS into the quasiparticle continuum at $E=\pm\Delta_0$ at some
finite values of $\phi$ which we denote $\phi_c$. Both features will lead to a sudden change in the slope,
$\partial E_n(\phi)/\partial \phi$. This, provided that the ABS is occupied, will leave fingerprints
in the current-phase relation at low temperatures.

\section{Current-phase relation}

After the ABS spectrum $\{E_n(\phi)\}$ is found for given $k_y$, the $k_y$-resolved
supercurrent is given by
the phase dispersion of $\{E_n\}$,
\begin{equation}
I(k_y,\phi)=\frac{2e}{\hbar}\displaystyle\sum_{n} \frac{\partial E_n}{\partial\phi}\frac{1}{e^{E_n/T}+1},
\end{equation}
where $T$ is the temperature, the Boltzmann constant $k_B$ is set to 1, and the sum is over all ABS energies
(continuum quasiparticle excitations give zero net contribution to the supercurrent). We can further exploit the particle-hole symmetry to rewrite this as
\begin{equation}
I(k_y,\phi)=-\frac{2e}{\hbar}\displaystyle\sum_{E_n\geq 0} \frac{\partial E_n}{\partial\phi}\tanh{\frac{E_n}{2T}}.
\end{equation}
 Numerically, we approximate the derivative as
\begin{equation}
 \frac{\partial E}{\partial\phi}=\frac{E(\phi+\epsilon)-E(\phi-\epsilon)}{2\epsilon},
\end{equation}
for small $\epsilon\ll 1$.

\begin{figure}[h]
\includegraphics[width=0.45\textwidth]{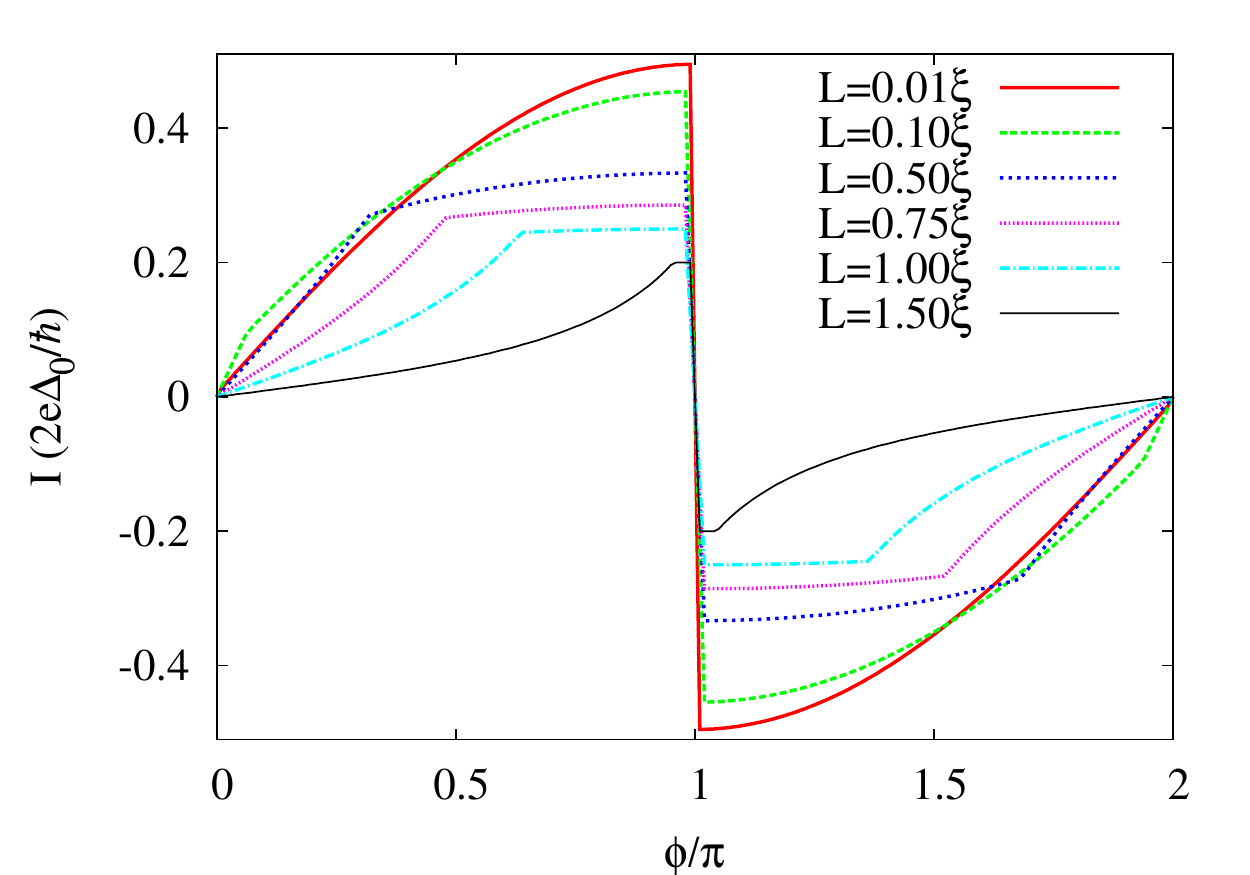}
\includegraphics[width=0.45\textwidth]{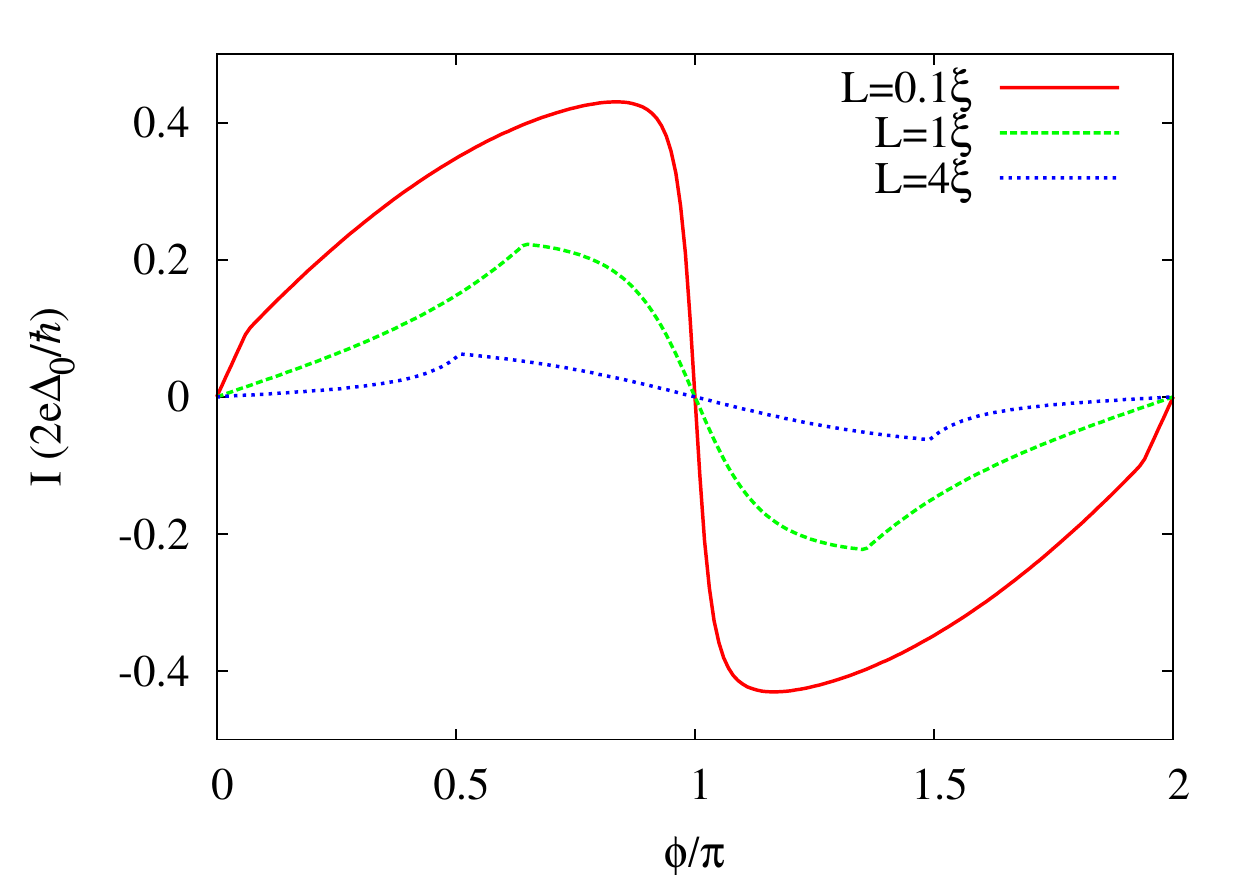}
\caption{Momentum ($k_y$) resolved supercurrent at zero temperature as function of $\phi$.
Top panel: $k_y=0$. Bottom panel: $k_y=0.2\xi^{-1}$. The junction parameters are the same as Fig. \ref{abs-plot}.
}\label{cpr-ky}
\end{figure}

Fig.~\ref{cpr-ky} shows the $I(k_y,\phi)$ corresponding to the ABS spectrum
shown in Fig. \ref{abs-plot}. For $k_y=0$, the sudden jump and sign change of $I$ 
at $\phi=\pi$ can be traced back to the zero energy crossings in the ABS. 
Due to the crossing, the slope of occupied ABS at zero temperature
experiences a sign change.
For a short junction, $I(k_y=0,\phi)=I_c \sin(\phi/2)$, in agreement of the analysis of Ref.~\onlinecite{fu_superconducting_2008}.
The remaining sharp turns, i.e., discontinuities in the derivatives of $I$, occur
at $\phi_c$ where the ABS reaches the superconducting gap and is absorbed into the 
quasiparticle continuum.
For finite $k_y$, the sudden drop at $\phi=\pi$ is replaced by a smooth variation, 
but the sign change of $I$ remains. Particularly, we observe that $I(\phi=\pi)=0$.

\begin{figure}
\includegraphics[width=0.45\textwidth]{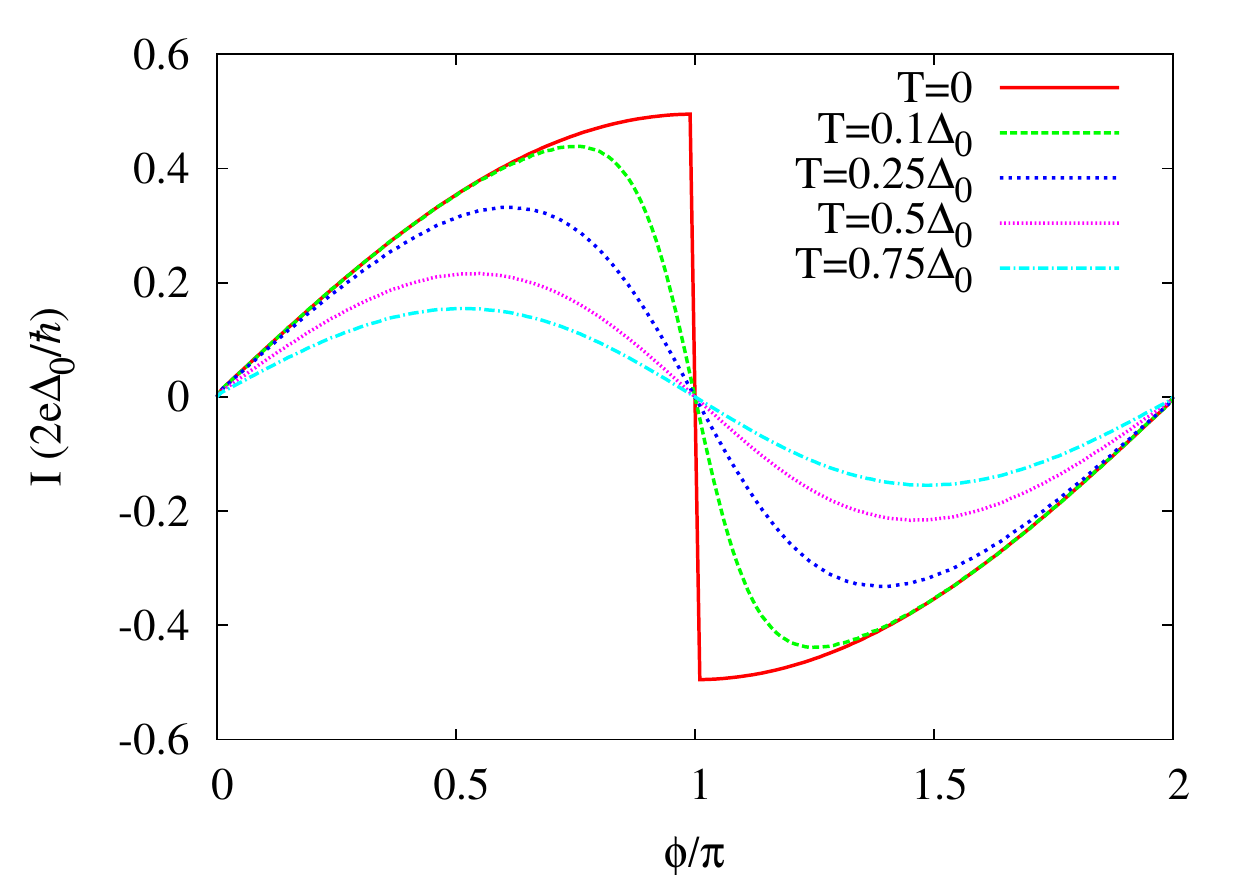}
\includegraphics[width=0.45\textwidth]{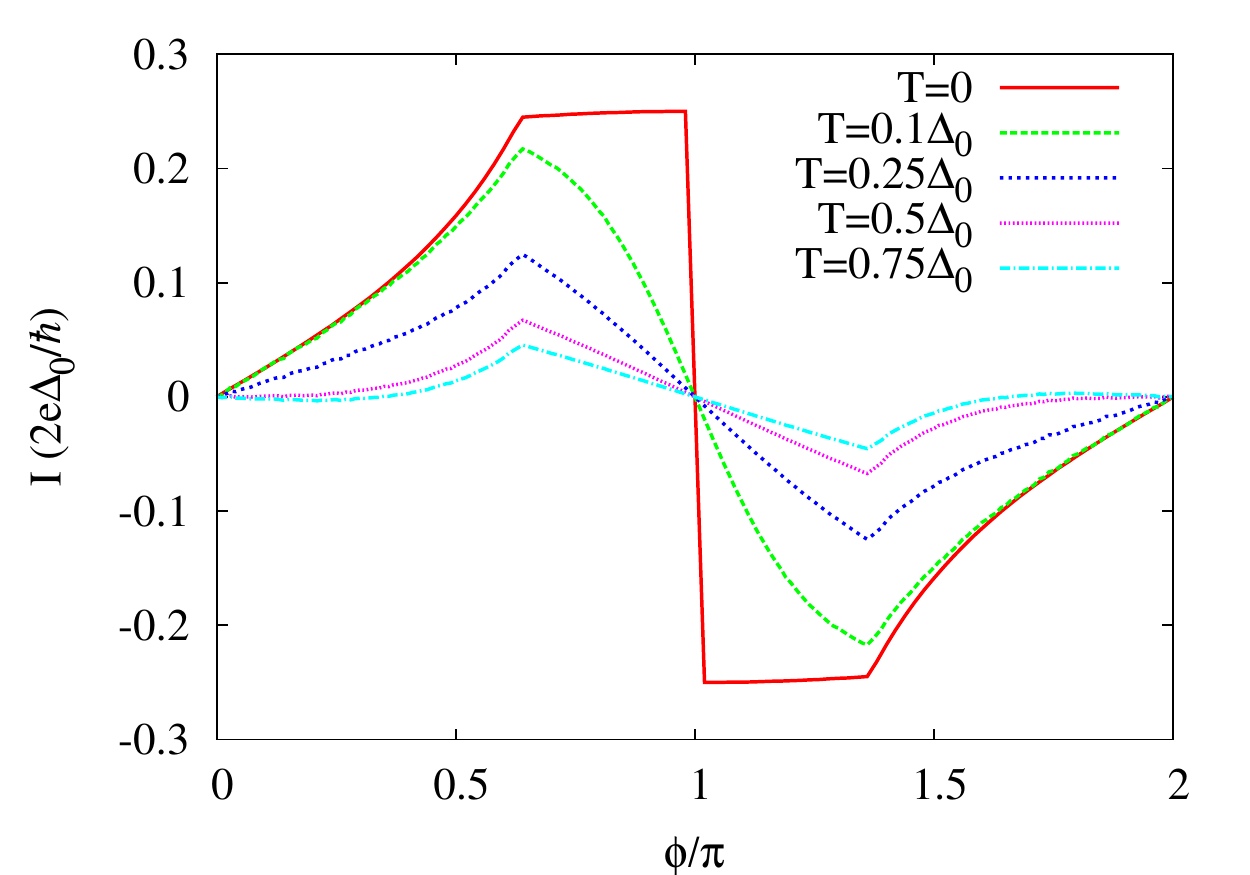}
\caption{Effect of temperature on $I(k_y=0,\phi)$. Upper panel: a short junction
with $L=0.01\xi$. Lower panel: a long junction with $L=\xi$. $\mu_S=2\Delta_0$, $\mu_M=0$, $W=\infty$.
}
\label{temp}
\end{figure}

Now we discuss the effect of temperature. Fig. \ref{temp} illustrates the evolution of 
$I(k_y=0,\phi)$ for a short and a long junction. As $T$ rises, the sawtooth-shaped CPR
gradually becomes a sine function at high temperatures. This is due
to the thermal population of all ABS levels which tends to smooth out the 
sudden jump at $\phi=\pi$.
By comparison we see that the sharp turns in $I(\phi)$ for long junctions survive even at
finite temperatures. These features are thus in principle observable in future experiments. 

\begin{figure}
\includegraphics[width=0.45\textwidth]{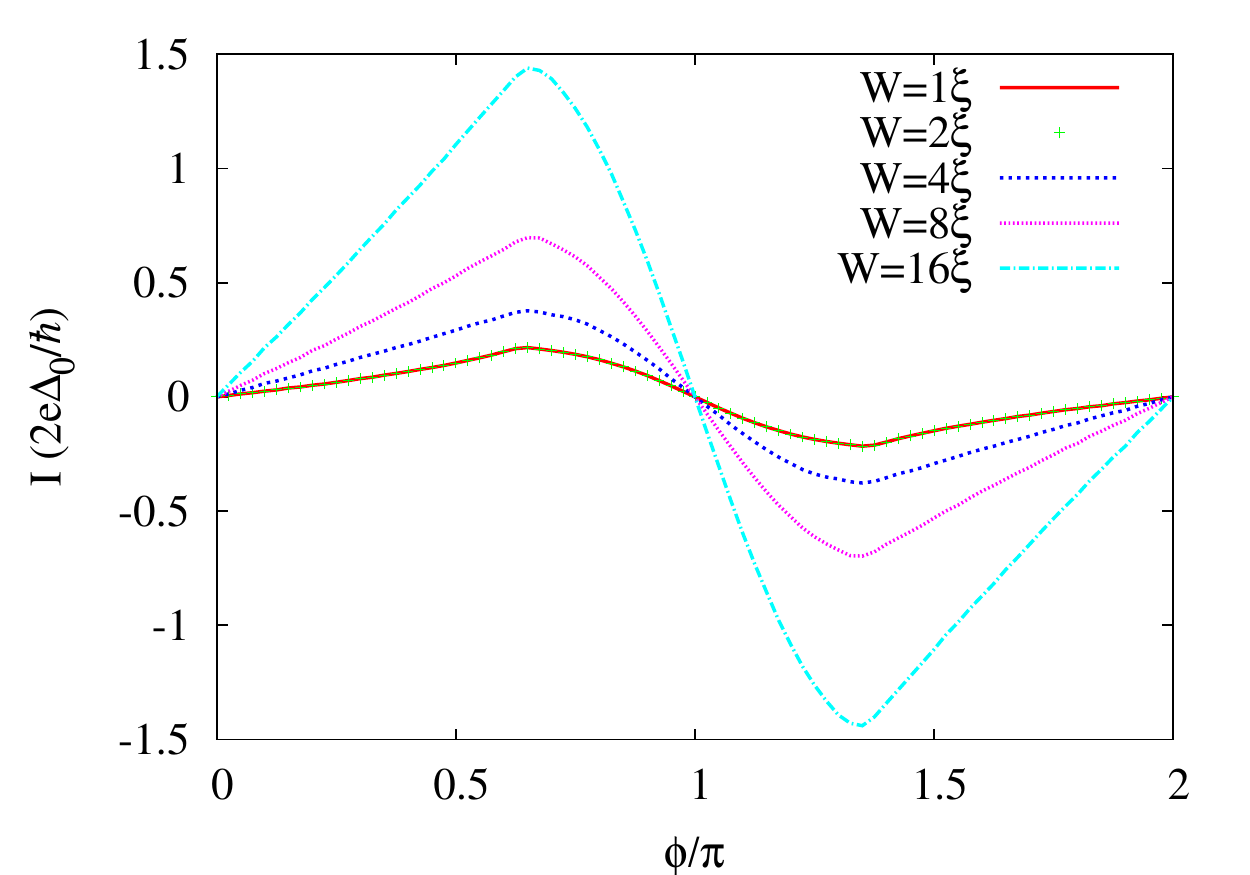}
\caption{Total supercurrent for junctions with finite width $W$.
$\mu_S=2\Delta_0$, $L=\xi$, $\mu_M=0$, $T=0.1\Delta_0$.}
\label{W}
\end{figure}

In order to get the total supercurrent, we must sum over all possible values of $k_y$,
\begin{equation}
I(\phi)=\int d k_y I(k_y,\phi).
\label{inte-I}
\end{equation}
For infinitely wide junctions, the integration goes from $-k_F^*$ to $k_F^*$, with $k_F^*=(\mu_{M}+\Delta_0)/v$.
For a junction with finite width $W$, we assume for simplicity open boundary
conditions at $y=0$ and $W$ (this may not accurately represent certain experiments, such as those in Ref.~\onlinecite{2012arXiv1202.2323W}). Then $k_y$ is quantized, $k_y^n=\pi n/W$, with $n$ being an integer,
and the integral in Eq.~(\ref{inte-I}) is replaced by a discrete sum over $|k_y^n|<k_F^*$. 
This yields the current phase relation. Fig.~\ref{W} compares the CPR 
of junctions with various widths. As the number of transverse modes increases,
the total current increases accordingly. The overall shape of the CPR remains
roughly the same, however.

\begin{figure}
\includegraphics[width=0.45\textwidth]{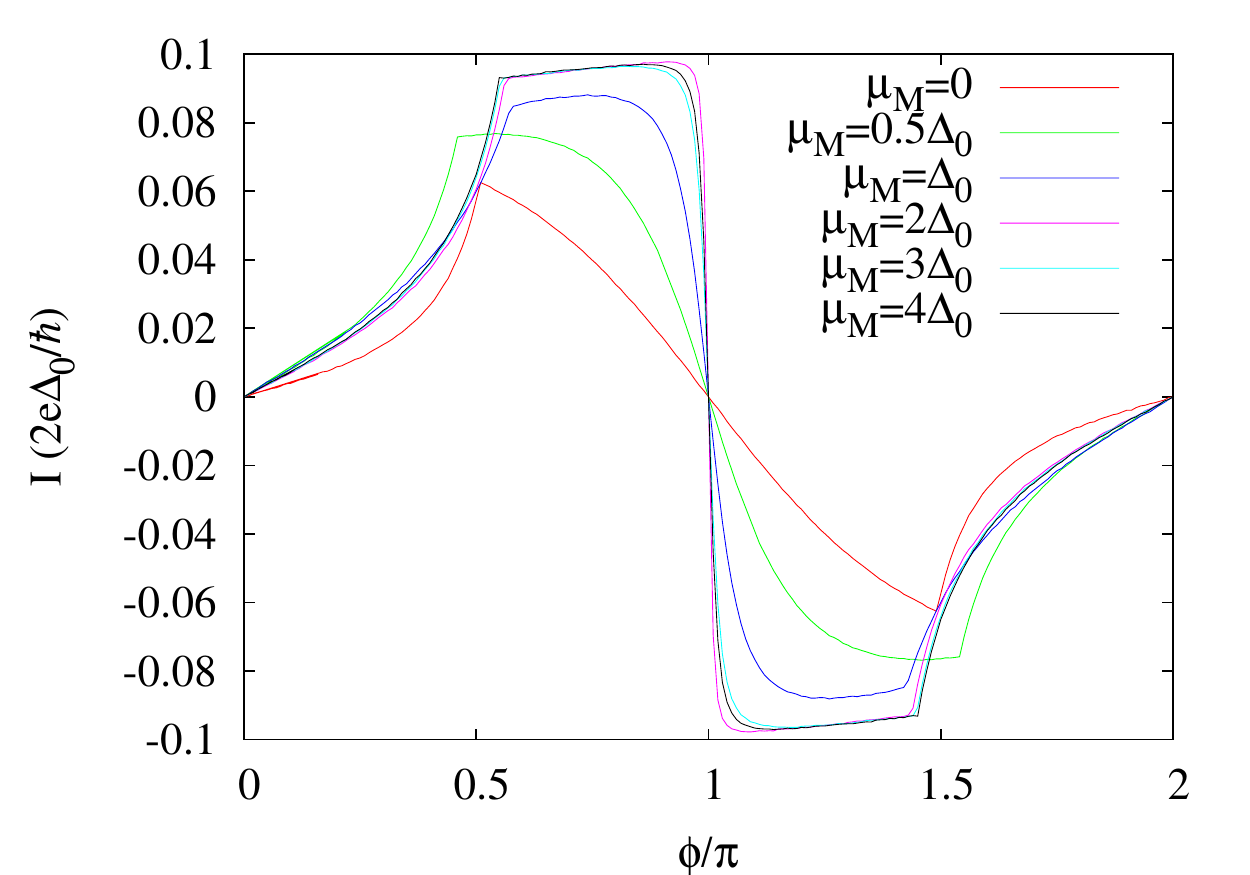}
\caption{Momentum ($k_y$) resolved supercurrent for junctions with various helical metal chemical potential $\mu_M$.
$\mu_S=2\Delta_0$, $k_y=0.2\xi^{-1}$, $L=\xi$, $T=0.1\Delta_0$.}
\label{muM:ky}
\end{figure}

\begin{figure}
\includegraphics[width=0.45\textwidth]{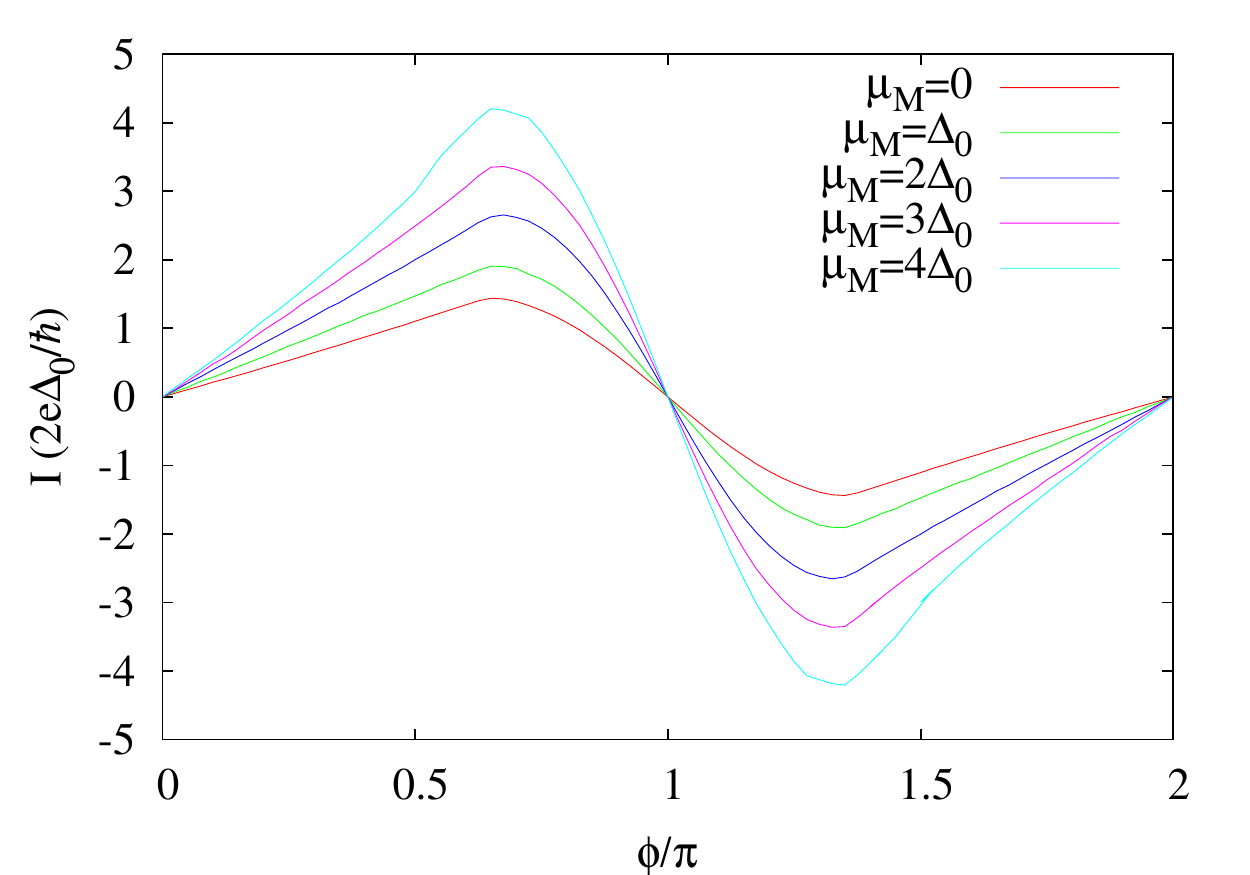}
\caption{Total supercurrent for junctions with various helical metal chemical potential $\mu_M$.
$\mu_S=2\Delta_0$, $L=\xi$, $T=0.1\Delta_0$, $W$=16$\xi$.}
\label{muM:total}
\end{figure}

Finally, we examine how the chemical potential of the helical metal, $\mu_M$, affects the CPR with $\mu_S$ fixed. For $k_y=0$, changing $\mu_M$ has no effect. This is explained by the lack of any $\mu_M$ dependance in the ABS spectrum we found analytically for $k_y=0$, as seen in Eq.~(\ref{transcen}). For $k_y \neq 0$, however, $\mu_M$ does affect the current.
As shown in Fig.~\ref{muM:ky}, the $k_y$-resolved current generally increases with $\mu_M$ until $\mu_M=\mu_S$, at which point it saturates and varies very little. The total current, as shown in Fig.~\ref{muM:total}, has a relatively consistent shape, but its magnitude increases with $\mu_M$ primarily because 
more transverse modes are available for large $\mu_M$.

\section{Scaling of the critical current with junction length}

\begin{figure}
\includegraphics[width=0.45\textwidth]{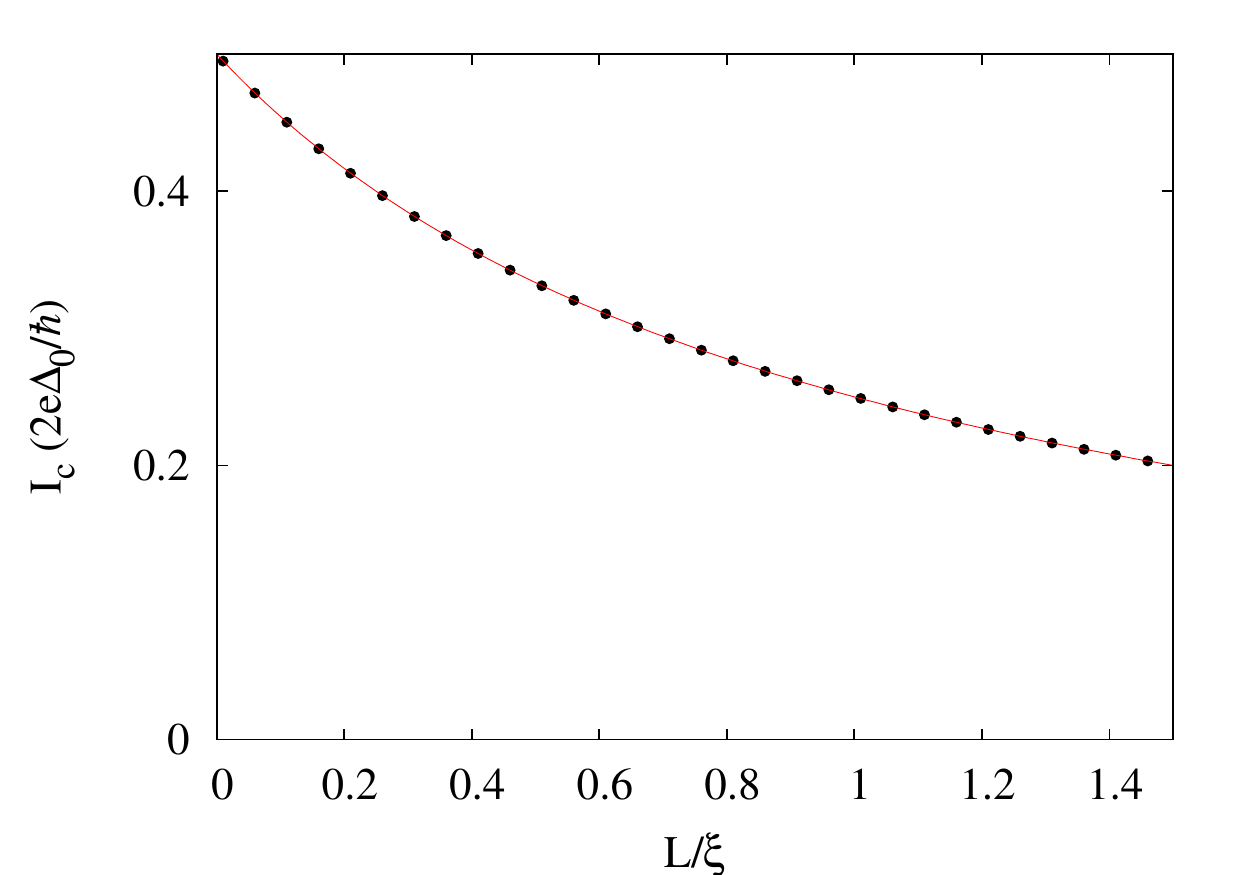}
\caption{The supercurrent at zero temperature through a single mode ($k_y=0$) 
TI nanoribbon of length $L$. $\mu_S=2\Delta_0$, $\mu_M=0$.}
\label{scaling}
\end{figure}

Now we consider the critical current $I_c$ for a junction through
a topological insulator nanoribbon of length $L$ with a single transverse mode $k_y=0$ (due to small $W$).
Fig.~\ref{scaling} shows the numerically calculated $I_c$ as a function of $L$. Since $\mu_M=0$,
there are no states (propagating modes) available at the Fermi level of the helical metal. Naively,
one would expect the critical current to decay rapidly with $L$ --- indeed, the decay should
be exponential in $L$ if the helical metal is replaced by a normal insulator. This interesting
problem was investigated for the Josephson effect through another semimetal, graphene, with chemical potential tuned right 
to the Dirac point (for review, see Ref.~\onlinecite{RevModPhys.80.1337}). There it was realized that graphene behaves very much like a disordered metal. 
$I_c$ reaches a minimum value for 
$\mu_M=0$, where $I_c$ scales with $1/L$. For helical metal, we find that the numerical data 
in Fig. \ref{scaling} can be fit by
\begin{equation}
 I_c=\frac{I_0}{L/\xi+1}.
 \label{scale-eq}
\end{equation}
Actually, since we have derived analytical results for the ABS spectrum for $k_y=0$, Eq.~(\ref{transcen}),
we can derive the equation above analytically. Taking the derivative of Eq.~(\ref{transcen}) with
respect to $\phi$ on both sides and recognizing that that the critical current corresponds to
$\phi$ approaching $\pi$, we obtain Eq.~(\ref{scale-eq}).

\section{Summary}
We have described how the current-phase relation of the Josephson effect through helical metal can be
obtained for general parameters, including the length and width of the junction, the chemical potentials,
and temperature, as motivated by recent experiments. 
A few useful analytical results that we derived, especially for single mode nanoribbon junctions,
 yield nontrivial predictions on the scaling of the critical current with the length of the junction.
The numerical algorithm outlined here can be implemented
to model the supercurrent flow in experiments where the current is entirely carried
by the surface of the helical metal. A detailed understanding of the spectra and CPR of these Josephson junctions
will contribute to the general goal of using them for superconducting devices and topological qubits.

\begin{acknowledgements}
We thank Mahmoud Lababidi for initial work on this project and very helpful discussions. 
This work is supported by NIST Grant No.~70NANB7H6138 Am 001 and ONR Grant No. N00014-09-1-1025A.
\end{acknowledgements}

\bibliographystyle{apsrev}
\bibliography{jj}{}

\begin{thebibliography}{35}
\expandafter\ifx\csname natexlab\endcsname\relax\def\natexlab#1{#1}\fi
\expandafter\ifx\csname bibnamefont\endcsname\relax
  \def\bibnamefont#1{#1}\fi
\expandafter\ifx\csname bibfnamefont\endcsname\relax
  \def\bibfnamefont#1{#1}\fi
\expandafter\ifx\csname citenamefont\endcsname\relax
  \def\citenamefont#1{#1}\fi
\expandafter\ifx\csname url\endcsname\relax
  \def\url#1{\texttt{#1}}\fi
\expandafter\ifx\csname urlprefix\endcsname\relax\def\urlprefix{URL }\fi
\providecommand{\bibinfo}[2]{#2}
\providecommand{\eprint}[2][]{\url{#2}}

\bibitem[{\citenamefont{Likharev}(1979)}]{RevModPhys.51.101}
\bibinfo{author}{\bibfnamefont{K.~K.} \bibnamefont{Likharev}},
  \bibinfo{journal}{Rev. Mod. Phys.} \textbf{\bibinfo{volume}{51}},
  \bibinfo{pages}{101} (\bibinfo{year}{1979}),
  \urlprefix\url{http://link.aps.org/doi/10.1103/RevModPhys.51.101}.

\bibitem[{\citenamefont{Makhlin et~al.}(2001)\citenamefont{Makhlin, Sch\"on,
  and Shnirman}}]{RevModPhys.73.357}
\bibinfo{author}{\bibfnamefont{Y.}~\bibnamefont{Makhlin}},
  \bibinfo{author}{\bibfnamefont{G.}~\bibnamefont{Sch\"on}}, \bibnamefont{and}
  \bibinfo{author}{\bibfnamefont{A.}~\bibnamefont{Shnirman}},
  \bibinfo{journal}{Rev. Mod. Phys.} \textbf{\bibinfo{volume}{73}},
  \bibinfo{pages}{357} (\bibinfo{year}{2001}),
  \urlprefix\url{http://link.aps.org/doi/10.1103/RevModPhys.73.357}.

\bibitem[{\citenamefont{Schoelkopf and Girvin}(2008)}]{Schoelkopf:2008fk}
\bibinfo{author}{\bibfnamefont{R.~J.} \bibnamefont{Schoelkopf}}
  \bibnamefont{and} \bibinfo{author}{\bibfnamefont{S.~M.}
  \bibnamefont{Girvin}}, \bibinfo{journal}{Nature}
  \textbf{\bibinfo{volume}{451}}, \bibinfo{pages}{664} (\bibinfo{year}{2008}),
  \urlprefix\url{http://dx.doi.org/10.1038/451664a}.

\bibitem[{\citenamefont{Fu and Kane}(2008)}]{fu_superconducting_2008}
\bibinfo{author}{\bibfnamefont{L.}~\bibnamefont{Fu}} \bibnamefont{and}
  \bibinfo{author}{\bibfnamefont{C.~L.} \bibnamefont{Kane}},
  \bibinfo{journal}{Physical Review Letters} \textbf{\bibinfo{volume}{100}},
  \bibinfo{pages}{096407} (\bibinfo{year}{2008}),
  \urlprefix\url{http://link.aps.org/doi/10.1103/PhysRevLett.100.096407}.

\bibitem[{\citenamefont{Jiang et~al.}(2011)\citenamefont{Jiang, Kane, and
  Preskill}}]{PhysRevLett.106.130504}
\bibinfo{author}{\bibfnamefont{L.}~\bibnamefont{Jiang}},
  \bibinfo{author}{\bibfnamefont{C.~L.} \bibnamefont{Kane}}, \bibnamefont{and}
  \bibinfo{author}{\bibfnamefont{J.}~\bibnamefont{Preskill}},
  \bibinfo{journal}{Phys. Rev. Lett.} \textbf{\bibinfo{volume}{106}},
  \bibinfo{pages}{130504} (\bibinfo{year}{2011}),
  \urlprefix\url{http://link.aps.org/doi/10.1103/PhysRevLett.106.130504}.

\bibitem[{\citenamefont{Bonderson and Lutchyn}(2011)}]{PhysRevLett.106.130505}
\bibinfo{author}{\bibfnamefont{P.}~\bibnamefont{Bonderson}} \bibnamefont{and}
  \bibinfo{author}{\bibfnamefont{R.~M.} \bibnamefont{Lutchyn}},
  \bibinfo{journal}{Phys. Rev. Lett.} \textbf{\bibinfo{volume}{106}},
  \bibinfo{pages}{130505} (\bibinfo{year}{2011}),
  \urlprefix\url{http://link.aps.org/doi/10.1103/PhysRevLett.106.130505}.

\bibitem[{\citenamefont{Golubov et~al.}(2004)\citenamefont{Golubov, Kupriyanov,
  and Il'ichev}}]{RevModPhys.76.411}
\bibinfo{author}{\bibfnamefont{A.~A.} \bibnamefont{Golubov}},
  \bibinfo{author}{\bibfnamefont{M.~Y.} \bibnamefont{Kupriyanov}},
  \bibnamefont{and} \bibinfo{author}{\bibfnamefont{E.}~\bibnamefont{Il'ichev}},
  \bibinfo{journal}{Rev. Mod. Phys.} \textbf{\bibinfo{volume}{76}},
  \bibinfo{pages}{411} (\bibinfo{year}{2004}),
  \urlprefix\url{http://link.aps.org/doi/10.1103/RevModPhys.76.411}.

\bibitem[{\citenamefont{Josephson}(1962)}]{Josephson1962251}
\bibinfo{author}{\bibfnamefont{B.}~\bibnamefont{Josephson}},
  \bibinfo{journal}{Physics Letters} \textbf{\bibinfo{volume}{1}},
  \bibinfo{pages}{251 } (\bibinfo{year}{1962}), ISSN \bibinfo{issn}{0031-9163},
  \urlprefix\url{http://www.sciencedirect.com/science/article/pii/003191636291%
3690}.

\bibitem[{\citenamefont{Kulik and Omelyanchuk}(1977)}]{ko}
\bibinfo{author}{\bibfnamefont{I.}~\bibnamefont{Kulik}} \bibnamefont{and}
  \bibinfo{author}{\bibfnamefont{A.}~\bibnamefont{Omelyanchuk}},
  \bibinfo{journal}{Sov. J. Low Temp. Phys.} \textbf{\bibinfo{volume}{3}},
  \bibinfo{pages}{459} (\bibinfo{year}{1977}).

\bibitem[{\citenamefont{Hasan and Kane}(2010)}]{hasan_colloquium:_2010}
\bibinfo{author}{\bibfnamefont{M.~Z.} \bibnamefont{Hasan}} \bibnamefont{and}
  \bibinfo{author}{\bibfnamefont{C.~L.} \bibnamefont{Kane}},
  \bibinfo{journal}{Reviews of Modern Physics} \textbf{\bibinfo{volume}{82}},
  \bibinfo{pages}{3045} (\bibinfo{year}{2010}),
  \urlprefix\url{http://link.aps.org/doi/10.1103/RevModPhys.82.3045}.

\bibitem[{\citenamefont{Qi and Zhang}(2011)}]{qi_topological_2011}
\bibinfo{author}{\bibfnamefont{X.}~\bibnamefont{Qi}} \bibnamefont{and}
  \bibinfo{author}{\bibfnamefont{S.}~\bibnamefont{Zhang}},
  \bibinfo{journal}{Reviews of Modern Physics} \textbf{\bibinfo{volume}{83}},
  \bibinfo{pages}{1057} (\bibinfo{year}{2011}),
  \urlprefix\url{http://link.aps.org/doi/10.1103/RevModPhys.83.1057}.

\bibitem[{\citenamefont{Beenakker}(2006)}]{PhysRevLett.97.067007}
\bibinfo{author}{\bibfnamefont{C.~W.~J.} \bibnamefont{Beenakker}},
  \bibinfo{journal}{Phys. Rev. Lett.} \textbf{\bibinfo{volume}{97}},
  \bibinfo{pages}{067007} (\bibinfo{year}{2006}),
  \urlprefix\url{http://link.aps.org/doi/10.1103/PhysRevLett.97.067007}.

\bibitem[{\citenamefont{Sac\'ep\'e et~al.}(2011)\citenamefont{Sac\'ep\'e,
  Oostinga, Li, Ubaldini, Couto, Giannini, and Morpurgo}}]{SacACpAC:2011vn}
\bibinfo{author}{\bibfnamefont{B.}~\bibnamefont{Sac\'ep\'e}},
  \bibinfo{author}{\bibfnamefont{J.~B.} \bibnamefont{Oostinga}},
  \bibinfo{author}{\bibfnamefont{J.}~\bibnamefont{Li}},
  \bibinfo{author}{\bibfnamefont{A.}~\bibnamefont{Ubaldini}},
  \bibinfo{author}{\bibfnamefont{N.~J.~G.} \bibnamefont{Couto}},
  \bibinfo{author}{\bibfnamefont{E.}~\bibnamefont{Giannini}}, \bibnamefont{and}
  \bibinfo{author}{\bibfnamefont{A.~F.} \bibnamefont{Morpurgo}},
  \bibinfo{journal}{Nat Commun} \textbf{\bibinfo{volume}{2}}
  (\bibinfo{year}{2011}), \urlprefix\url{http://dx.doi.org/10.1038/ncomms1586}.

\bibitem[{\citenamefont{Zhang et~al.}(2011{\natexlab{a}})\citenamefont{Zhang,
  Wang, DaSilva, Lee, Gutierrez, Chan, Jain, and Samarth}}]{PhysRevB.84.165120}
\bibinfo{author}{\bibfnamefont{D.}~\bibnamefont{Zhang}},
  \bibinfo{author}{\bibfnamefont{J.}~\bibnamefont{Wang}},
  \bibinfo{author}{\bibfnamefont{A.~M.} \bibnamefont{DaSilva}},
  \bibinfo{author}{\bibfnamefont{J.~S.} \bibnamefont{Lee}},
  \bibinfo{author}{\bibfnamefont{H.~R.} \bibnamefont{Gutierrez}},
  \bibinfo{author}{\bibfnamefont{M.~H.~W.} \bibnamefont{Chan}},
  \bibinfo{author}{\bibfnamefont{J.}~\bibnamefont{Jain}}, \bibnamefont{and}
  \bibinfo{author}{\bibfnamefont{N.}~\bibnamefont{Samarth}},
  \bibinfo{journal}{Phys. Rev. B} \textbf{\bibinfo{volume}{84}},
  \bibinfo{pages}{165120} (\bibinfo{year}{2011}{\natexlab{a}}),
  \urlprefix\url{http://link.aps.org/doi/10.1103/PhysRevB.84.165120}.

\bibitem[{\citenamefont{Veldhorst et~al.}(2012)\citenamefont{Veldhorst,
  Snelder, Hoek, Gang, Guduru, Wang, Zeitler, van~der Wiel, Golubov, Hilgenkamp
  et~al.}}]{Veldhorst:2012uq}
\bibinfo{author}{\bibfnamefont{M.}~\bibnamefont{Veldhorst}},
  \bibinfo{author}{\bibfnamefont{M.}~\bibnamefont{Snelder}},
  \bibinfo{author}{\bibfnamefont{M.}~\bibnamefont{Hoek}},
  \bibinfo{author}{\bibfnamefont{T.}~\bibnamefont{Gang}},
  \bibinfo{author}{\bibfnamefont{V.~K.} \bibnamefont{Guduru}},
  \bibinfo{author}{\bibfnamefont{X.~L.} \bibnamefont{Wang}},
  \bibinfo{author}{\bibfnamefont{U.}~\bibnamefont{Zeitler}},
  \bibinfo{author}{\bibfnamefont{W.~G.} \bibnamefont{van~der Wiel}},
  \bibinfo{author}{\bibfnamefont{A.~A.} \bibnamefont{Golubov}},
  \bibinfo{author}{\bibfnamefont{H.}~\bibnamefont{Hilgenkamp}},
  \bibnamefont{et~al.}, \bibinfo{journal}{Nat Mater}
  \textbf{\bibinfo{volume}{11}}, \bibinfo{pages}{417} (\bibinfo{year}{2012}),
  \urlprefix\url{http://dx.doi.org/10.1038/nmat3255}.

\bibitem[{\citenamefont{Qu et~al.}(2012)\citenamefont{Qu, Yang, Shen, Ding,
  Chen, Ji, Liu, Fan, Jing, Yang et~al.}}]{Qu:2012kx}
\bibinfo{author}{\bibfnamefont{F.}~\bibnamefont{Qu}},
  \bibinfo{author}{\bibfnamefont{F.}~\bibnamefont{Yang}},
  \bibinfo{author}{\bibfnamefont{J.}~\bibnamefont{Shen}},
  \bibinfo{author}{\bibfnamefont{Y.}~\bibnamefont{Ding}},
  \bibinfo{author}{\bibfnamefont{J.}~\bibnamefont{Chen}},
  \bibinfo{author}{\bibfnamefont{Z.}~\bibnamefont{Ji}},
  \bibinfo{author}{\bibfnamefont{G.}~\bibnamefont{Liu}},
  \bibinfo{author}{\bibfnamefont{J.}~\bibnamefont{Fan}},
  \bibinfo{author}{\bibfnamefont{X.}~\bibnamefont{Jing}},
  \bibinfo{author}{\bibfnamefont{C.}~\bibnamefont{Yang}}, \bibnamefont{et~al.},
  \bibinfo{journal}{Sci. Rep.} \textbf{\bibinfo{volume}{2}}
  (\bibinfo{year}{2012}), \urlprefix\url{http://dx.doi.org/10.1038/srep00339}.

\bibitem[{\citenamefont{{Williams} et~al.}(2012)\citenamefont{{Williams},
  {Bestwick}, {Gallagher}, {Hong}, {Cui}, {Bleich}, {Analytis}, {Fisher}, and
  {Goldhaber-Gordon}}}]{2012arXiv1202.2323W}
\bibinfo{author}{\bibfnamefont{J.~R.} \bibnamefont{{Williams}}},
  \bibinfo{author}{\bibfnamefont{A.~J.} \bibnamefont{{Bestwick}}},
  \bibinfo{author}{\bibfnamefont{P.}~\bibnamefont{{Gallagher}}},
  \bibinfo{author}{\bibfnamefont{S.~S.} \bibnamefont{{Hong}}},
  \bibinfo{author}{\bibfnamefont{Y.}~\bibnamefont{{Cui}}},
  \bibinfo{author}{\bibfnamefont{A.~S.} \bibnamefont{{Bleich}}},
  \bibinfo{author}{\bibfnamefont{J.~G.} \bibnamefont{{Analytis}}},
  \bibinfo{author}{\bibfnamefont{I.~R.} \bibnamefont{{Fisher}}},
  \bibnamefont{and}
  \bibinfo{author}{\bibfnamefont{D.}~\bibnamefont{{Goldhaber-Gordon}}},
  \bibinfo{journal}{ArXiv e-prints}  (\bibinfo{year}{2012}),
  \eprint{1202.2323}.

\bibitem[{\citenamefont{{Yang} et~al.}(2012)\citenamefont{{Yang}, {Qu}, {Shen},
  {Ding}, {Chen}, {Ji}, {Liu}, {Fan}, {Yang}, {Fu}
  et~al.}}]{2012arXiv1206.6178Y}
\bibinfo{author}{\bibfnamefont{F.}~\bibnamefont{{Yang}}},
  \bibinfo{author}{\bibfnamefont{F.}~\bibnamefont{{Qu}}},
  \bibinfo{author}{\bibfnamefont{J.}~\bibnamefont{{Shen}}},
  \bibinfo{author}{\bibfnamefont{Y.}~\bibnamefont{{Ding}}},
  \bibinfo{author}{\bibfnamefont{J.}~\bibnamefont{{Chen}}},
  \bibinfo{author}{\bibfnamefont{Z.}~\bibnamefont{{Ji}}},
  \bibinfo{author}{\bibfnamefont{G.}~\bibnamefont{{Liu}}},
  \bibinfo{author}{\bibfnamefont{J.}~\bibnamefont{{Fan}}},
  \bibinfo{author}{\bibfnamefont{C.}~\bibnamefont{{Yang}}},
  \bibinfo{author}{\bibfnamefont{L.}~\bibnamefont{{Fu}}}, \bibnamefont{et~al.},
  \bibinfo{journal}{ArXiv e-prints}  (\bibinfo{year}{2012}),
  \eprint{1206.6178}.

\bibitem[{\citenamefont{Checkelsky et~al.}(2011)\citenamefont{Checkelsky, Hor,
  Cava, and Ong}}]{PhysRevLett.106.196801}
\bibinfo{author}{\bibfnamefont{J.~G.} \bibnamefont{Checkelsky}},
  \bibinfo{author}{\bibfnamefont{Y.~S.} \bibnamefont{Hor}},
  \bibinfo{author}{\bibfnamefont{R.~J.} \bibnamefont{Cava}}, \bibnamefont{and}
  \bibinfo{author}{\bibfnamefont{N.~P.} \bibnamefont{Ong}},
  \bibinfo{journal}{Phys. Rev. Lett.} \textbf{\bibinfo{volume}{106}},
  \bibinfo{pages}{196801} (\bibinfo{year}{2011}),
  \urlprefix\url{http://link.aps.org/doi/10.1103/PhysRevLett.106.196801}.

\bibitem[{\citenamefont{Wang et~al.}(2012)\citenamefont{Wang, Xiu, Cheng, He,
  Lang, Tang, Kou, Yu, Jiang, Chen et~al.}}]{doi:10.1021/nl202920p}
\bibinfo{author}{\bibfnamefont{Y.}~\bibnamefont{Wang}},
  \bibinfo{author}{\bibfnamefont{F.}~\bibnamefont{Xiu}},
  \bibinfo{author}{\bibfnamefont{L.}~\bibnamefont{Cheng}},
  \bibinfo{author}{\bibfnamefont{L.}~\bibnamefont{He}},
  \bibinfo{author}{\bibfnamefont{M.}~\bibnamefont{Lang}},
  \bibinfo{author}{\bibfnamefont{J.}~\bibnamefont{Tang}},
  \bibinfo{author}{\bibfnamefont{X.}~\bibnamefont{Kou}},
  \bibinfo{author}{\bibfnamefont{X.}~\bibnamefont{Yu}},
  \bibinfo{author}{\bibfnamefont{X.}~\bibnamefont{Jiang}},
  \bibinfo{author}{\bibfnamefont{Z.}~\bibnamefont{Chen}}, \bibnamefont{et~al.},
  \bibinfo{journal}{Nano Letters} \textbf{\bibinfo{volume}{12}},
  \bibinfo{pages}{1170} (\bibinfo{year}{2012}),
  \eprint{http://pubs.acs.org/doi/pdf/10.1021/nl202920p},
  \urlprefix\url{http://pubs.acs.org/doi/abs/10.1021/nl202920p}.

\bibitem[{\citenamefont{Kim et~al.}(2012)\citenamefont{Kim, Cho, Butch, Syers,
  Kirshenbaum, Adam, Paglione, and Fuhrer}}]{Kim:2012kx}
\bibinfo{author}{\bibfnamefont{D.}~\bibnamefont{Kim}},
  \bibinfo{author}{\bibfnamefont{S.}~\bibnamefont{Cho}},
  \bibinfo{author}{\bibfnamefont{N.~P.} \bibnamefont{Butch}},
  \bibinfo{author}{\bibfnamefont{P.}~\bibnamefont{Syers}},
  \bibinfo{author}{\bibfnamefont{K.}~\bibnamefont{Kirshenbaum}},
  \bibinfo{author}{\bibfnamefont{S.}~\bibnamefont{Adam}},
  \bibinfo{author}{\bibfnamefont{J.}~\bibnamefont{Paglione}}, \bibnamefont{and}
  \bibinfo{author}{\bibfnamefont{M.~S.} \bibnamefont{Fuhrer}},
  \bibinfo{journal}{Nat Phys} \textbf{\bibinfo{volume}{8}},
  \bibinfo{pages}{460} (\bibinfo{year}{2012}),
  \urlprefix\url{http://dx.doi.org/10.1038/nphys2286}.

\bibitem[{\citenamefont{Hong et~al.}(2012)\citenamefont{Hong, Cha, Kong, and
  Cui}}]{Hong:2012uq}
\bibinfo{author}{\bibfnamefont{S.~S.} \bibnamefont{Hong}},
  \bibinfo{author}{\bibfnamefont{J.~J.} \bibnamefont{Cha}},
  \bibinfo{author}{\bibfnamefont{D.}~\bibnamefont{Kong}}, \bibnamefont{and}
  \bibinfo{author}{\bibfnamefont{Y.}~\bibnamefont{Cui}}, \bibinfo{journal}{Nat
  Commun} \textbf{\bibinfo{volume}{3}} (\bibinfo{year}{2012}),
  \urlprefix\url{http://dx.doi.org/10.1038/ncomms1771}.

\bibitem[{\citenamefont{Zhang et~al.}(2011{\natexlab{b}})\citenamefont{Zhang,
  Chang, Zhang, Wen, Feng, Li, Liu, He, Wang, Chen et~al.}}]{Zhang:2011vn}
\bibinfo{author}{\bibfnamefont{J.}~\bibnamefont{Zhang}},
  \bibinfo{author}{\bibfnamefont{C.-Z.} \bibnamefont{Chang}},
  \bibinfo{author}{\bibfnamefont{Z.}~\bibnamefont{Zhang}},
  \bibinfo{author}{\bibfnamefont{J.}~\bibnamefont{Wen}},
  \bibinfo{author}{\bibfnamefont{X.}~\bibnamefont{Feng}},
  \bibinfo{author}{\bibfnamefont{K.}~\bibnamefont{Li}},
  \bibinfo{author}{\bibfnamefont{M.}~\bibnamefont{Liu}},
  \bibinfo{author}{\bibfnamefont{K.}~\bibnamefont{He}},
  \bibinfo{author}{\bibfnamefont{L.}~\bibnamefont{Wang}},
  \bibinfo{author}{\bibfnamefont{X.}~\bibnamefont{Chen}}, \bibnamefont{et~al.},
  \bibinfo{journal}{Nat Commun} \textbf{\bibinfo{volume}{2}}
  (\bibinfo{year}{2011}{\natexlab{b}}),
  \urlprefix\url{http://dx.doi.org/10.1038/ncomms1588}.

\bibitem[{\citenamefont{Kong et~al.}(2011)\citenamefont{Kong, Chen, Cha, Zhang,
  Analytis, Lai, Liu, Hong, Koski, Mo et~al.}}]{Kong:2011zr}
\bibinfo{author}{\bibfnamefont{D.}~\bibnamefont{Kong}},
  \bibinfo{author}{\bibfnamefont{Y.}~\bibnamefont{Chen}},
  \bibinfo{author}{\bibfnamefont{J.~J.} \bibnamefont{Cha}},
  \bibinfo{author}{\bibfnamefont{Q.}~\bibnamefont{Zhang}},
  \bibinfo{author}{\bibfnamefont{J.~G.} \bibnamefont{Analytis}},
  \bibinfo{author}{\bibfnamefont{K.}~\bibnamefont{Lai}},
  \bibinfo{author}{\bibfnamefont{Z.}~\bibnamefont{Liu}},
  \bibinfo{author}{\bibfnamefont{S.~S.} \bibnamefont{Hong}},
  \bibinfo{author}{\bibfnamefont{K.~J.} \bibnamefont{Koski}},
  \bibinfo{author}{\bibfnamefont{S.-K.} \bibnamefont{Mo}},
  \bibnamefont{et~al.}, \bibinfo{journal}{Nat Nano}
  \textbf{\bibinfo{volume}{6}}, \bibinfo{pages}{705} (\bibinfo{year}{2011}),
  \urlprefix\url{http://dx.doi.org/10.1038/nnano.2011.172}.

\bibitem[{\citenamefont{Arakane et~al.}(2012)\citenamefont{Arakane, Sato,
  Souma, Kosaka, Nakayama, Komatsu, Takahashi, Ren, Segawa, and
  Ando}}]{Arakane:2012ys}
\bibinfo{author}{\bibfnamefont{T.}~\bibnamefont{Arakane}},
  \bibinfo{author}{\bibfnamefont{T.}~\bibnamefont{Sato}},
  \bibinfo{author}{\bibfnamefont{S.}~\bibnamefont{Souma}},
  \bibinfo{author}{\bibfnamefont{K.}~\bibnamefont{Kosaka}},
  \bibinfo{author}{\bibfnamefont{K.}~\bibnamefont{Nakayama}},
  \bibinfo{author}{\bibfnamefont{M.}~\bibnamefont{Komatsu}},
  \bibinfo{author}{\bibfnamefont{T.}~\bibnamefont{Takahashi}},
  \bibinfo{author}{\bibfnamefont{Z.}~\bibnamefont{Ren}},
  \bibinfo{author}{\bibfnamefont{K.}~\bibnamefont{Segawa}}, \bibnamefont{and}
  \bibinfo{author}{\bibfnamefont{Y.}~\bibnamefont{Ando}}, \bibinfo{journal}{Nat
  Commun} \textbf{\bibinfo{volume}{3}} (\bibinfo{year}{2012}),
  \urlprefix\url{http://dx.doi.org/10.1038/ncomms1639}.

\bibitem[{\citenamefont{Tanaka et~al.}(2009)\citenamefont{Tanaka, Yokoyama, and
  Nagaosa}}]{PhysRevLett.103.107002}
\bibinfo{author}{\bibfnamefont{Y.}~\bibnamefont{Tanaka}},
  \bibinfo{author}{\bibfnamefont{T.}~\bibnamefont{Yokoyama}}, \bibnamefont{and}
  \bibinfo{author}{\bibfnamefont{N.}~\bibnamefont{Nagaosa}},
  \bibinfo{journal}{Phys. Rev. Lett.} \textbf{\bibinfo{volume}{103}},
  \bibinfo{pages}{107002} (\bibinfo{year}{2009}),
  \urlprefix\url{http://link.aps.org/doi/10.1103/PhysRevLett.103.107002}.

\bibitem[{\citenamefont{Linder et~al.}(2010)\citenamefont{Linder, Tanaka,
  Yokoyama, Sudb\o{}, and Nagaosa}}]{PhysRevB.81.184525}
\bibinfo{author}{\bibfnamefont{J.}~\bibnamefont{Linder}},
  \bibinfo{author}{\bibfnamefont{Y.}~\bibnamefont{Tanaka}},
  \bibinfo{author}{\bibfnamefont{T.}~\bibnamefont{Yokoyama}},
  \bibinfo{author}{\bibfnamefont{A.}~\bibnamefont{Sudb\o{}}}, \bibnamefont{and}
  \bibinfo{author}{\bibfnamefont{N.}~\bibnamefont{Nagaosa}},
  \bibinfo{journal}{Phys. Rev. B} \textbf{\bibinfo{volume}{81}},
  \bibinfo{pages}{184525} (\bibinfo{year}{2010}),
  \urlprefix\url{http://link.aps.org/doi/10.1103/PhysRevB.81.184525}.

\bibitem[{\citenamefont{{Yokoyama}}(2012)}]{2012arXiv1206.3831Y}
\bibinfo{author}{\bibfnamefont{T.}~\bibnamefont{{Yokoyama}}},
  \bibinfo{journal}{ArXiv e-prints}  (\bibinfo{year}{2012}),
  \eprint{1206.3831}.

\bibitem[{\citenamefont{Ioselevich and
  Feigel'man}(2011)}]{PhysRevLett.106.077003}
\bibinfo{author}{\bibfnamefont{P.~A.} \bibnamefont{Ioselevich}}
  \bibnamefont{and} \bibinfo{author}{\bibfnamefont{M.~V.}
  \bibnamefont{Feigel'man}}, \bibinfo{journal}{Phys. Rev. Lett.}
  \textbf{\bibinfo{volume}{106}}, \bibinfo{pages}{077003}
  (\bibinfo{year}{2011}),
  \urlprefix\url{http://link.aps.org/doi/10.1103/PhysRevLett.106.077003}.

\bibitem[{\citenamefont{Stanescu et~al.}(2010)\citenamefont{Stanescu, Sau,
  Lutchyn, and Das~Sarma}}]{PhysRevB.81.241310}
\bibinfo{author}{\bibfnamefont{T.~D.} \bibnamefont{Stanescu}},
  \bibinfo{author}{\bibfnamefont{J.~D.} \bibnamefont{Sau}},
  \bibinfo{author}{\bibfnamefont{R.~M.} \bibnamefont{Lutchyn}},
  \bibnamefont{and}
  \bibinfo{author}{\bibfnamefont{S.}~\bibnamefont{Das~Sarma}},
  \bibinfo{journal}{Phys. Rev. B} \textbf{\bibinfo{volume}{81}},
  \bibinfo{pages}{241310} (\bibinfo{year}{2010}),
  \urlprefix\url{http://link.aps.org/doi/10.1103/PhysRevB.81.241310}.

\bibitem[{\citenamefont{Lababidi and Zhao}(2011)}]{PhysRevB.83.184511}
\bibinfo{author}{\bibfnamefont{M.}~\bibnamefont{Lababidi}} \bibnamefont{and}
  \bibinfo{author}{\bibfnamefont{E.}~\bibnamefont{Zhao}},
  \bibinfo{journal}{Phys. Rev. B} \textbf{\bibinfo{volume}{83}},
  \bibinfo{pages}{184511} (\bibinfo{year}{2011}),
  \urlprefix\url{http://link.aps.org/doi/10.1103/PhysRevB.83.184511}.

\bibitem[{\citenamefont{{Grein} et~al.}(2011)\citenamefont{{Grein},
  {Michelsen}, and {Eschrig}}}]{2011arXiv1111.0445G}
\bibinfo{author}{\bibfnamefont{R.}~\bibnamefont{{Grein}}},
  \bibinfo{author}{\bibfnamefont{J.}~\bibnamefont{{Michelsen}}},
  \bibnamefont{and}
  \bibinfo{author}{\bibfnamefont{M.}~\bibnamefont{{Eschrig}}},
  \bibinfo{journal}{ArXiv e-prints}  (\bibinfo{year}{2011}),
  \eprint{1111.0445}.

\bibitem[{\citenamefont{Zhao et~al.}(2010)\citenamefont{Zhao, Zhang, and
  Lababidi}}]{PhysRevB.82.205331}
\bibinfo{author}{\bibfnamefont{E.}~\bibnamefont{Zhao}},
  \bibinfo{author}{\bibfnamefont{C.}~\bibnamefont{Zhang}}, \bibnamefont{and}
  \bibinfo{author}{\bibfnamefont{M.}~\bibnamefont{Lababidi}},
  \bibinfo{journal}{Phys. Rev. B} \textbf{\bibinfo{volume}{82}},
  \bibinfo{pages}{205331} (\bibinfo{year}{2010}),
  \urlprefix\url{http://link.aps.org/doi/10.1103/PhysRevB.82.205331}.

\bibitem[{\citenamefont{Press et~al.}(1992)\citenamefont{Press, Teukolsky,
  Vetterling, and Flannery}}]{nr}
\bibinfo{author}{\bibfnamefont{W.~H.} \bibnamefont{Press}},
  \bibinfo{author}{\bibfnamefont{S.~A.} \bibnamefont{Teukolsky}},
  \bibinfo{author}{\bibfnamefont{W.~T.} \bibnamefont{Vetterling}},
  \bibnamefont{and} \bibinfo{author}{\bibfnamefont{B.~P.}
  \bibnamefont{Flannery}}, \emph{\bibinfo{title}{Numerical Recipes in C: The
  Art of Scientific Computing}} (\bibinfo{publisher}{Cambridge University
  Press}, \bibinfo{year}{1992}), \bibinfo{edition}{2nd} ed.

\bibitem[{\citenamefont{Beenakker}(2008)}]{RevModPhys.80.1337}
\bibinfo{author}{\bibfnamefont{C.~W.~J.} \bibnamefont{Beenakker}},
  \bibinfo{journal}{Rev. Mod. Phys.} \textbf{\bibinfo{volume}{80}},
  \bibinfo{pages}{1337} (\bibinfo{year}{2008}),
  \urlprefix\url{http://link.aps.org/doi/10.1103/RevModPhys.80.1337}.

\end{thebibliography}
\end{document}